\documentclass[prl,aps]{revtex4}     
\usepackage{graphicx}
\usepackage{amsmath,dsfont}
\def\beq{\begin{equation}}
\def\eeq{\end{equation}}
\def\beqnn{\begin{displaymath}}
\def\eeqnn{\end{displaymath}}
\def\bea{\begin{eqnarray}}
\def\eea{\end{eqnarray}}

\def\eno#1{Eq.~(\ref{#1})}

\def\fno#1{Fig.~\ref{#1}}

\def\tst{\textstyle}

\def\al{\alpha}

\def\gam{\gamma}
\def\dta{\delta}

\def\tta{\theta}

\def\sig{\sigma}

\def\bap{{\bar p}}
\def\bas{{\bar s}}

\def\ptl{\partial}

\def\part#1#2{\frac{\ptl#1}{\ptl#2}}

\def\dint{\int\!\!\int}

\def\Tr{{\rm Tr\,}}

\def\gtwid{\mathrel{\raise.3ex\hbox{$>$\kern-.78em\lower1ex\hbox{$\sim$}}}}
\def\ltwid{\mathrel{\raise.3ex\hbox{$<$\kern-.78em\lower1ex\hbox{$\sim$}}}}

\def\hf{\frac{1}{2}}

\def\tshf{{\tst\hf}}

\def\ylm{Y_{\ell m}}
\def\Ylm{{{\cal Y}}_{\ell m}}

\def\tily{{\tilde Y}}

\def\typ0{\tily_{\ell+1,m}}

\def\tym0{\tily_{\ell-1,m}}

\def\Plp1{P_{\ell + 1}}
\def\Plm1{P_{\ell - 1}}

\def\llp1{\ell(\ell + 1)}
\def\tlp1{2\ell + 1}

\def\bJ{{\bf J}}

\def\ity{{\it y\/}\ }

\def\ahat{{\bf{\hat a}}}
\def\bhat{{\bf{\hat b}}}

\def\nhat{{\bf{\hat n}}}

\def\zhat{{\bf{\hat z}}}

\def\avg#1{\langle#1\rangle}

\def\ket#1{|#1\rangle}
\def\bra#1{\langle#1|}

\def\olap#1#2{\langle#1|#2\rangle}

\begin{document}

\title{Agnostic Detector Error, Wigner Functions, and the Classical Limit of the High-Spin
Einstein-Podolsky-Rosen Experiment}
\author{Anupam Garg}
\affiliation{Department of Physics and Astronomy, Northwestern University,
Evanston, Illinois 60208}

\date{\today}

\begin{abstract}
The spin-$j$ Einstein-Podolsky-Rosen experiment is studied with a view to understanding how classical
behaviour emerges as $j \to \infty$.
It is proposed that it is necessary to include detector error, which if it is be to viewed as
an essential aspect of the emergence of classicality, should be both minimal, i.e., no more than
necessary to wash out quantum mechanical behaviour, and agnostic, by which is meant that one should
be able to ascribe it to error in the preparation of the state just as well as to the detector.
Errors in the state preparation are discussed via the spin Wigner function.
An agnostic error protocol is described which appears to be minimal.
\end{abstract}

\maketitle

\section{Introduction}
\label{intro}

The suggestion that imperfections in the measurement process are necessary in order to
understand the connection between quantum and classical mechanics is surely a very old one,
and it is hard to pinpoint its exact genesis. It has been made by many authors with varying
degrees of emphasis and nuance, and with varying motivations, and we could not possibly know
of them all. Some references of which we are aware and which appear relevant to this paper
are\,\cite{reichenbach,ms82,kb07,lun09}. Our point of view is closest to Kofler
and Brukner\,\cite{kb07}, but goes well beyond it.

That some imperfection is required is suggested even by the elementary example of the
quantum mechanical probability distribution for position in a high-quantum-number
energy eigenstate of the simple harmonic oscillator\,\cite{shankar}. If we compare this to the 
classical distribution, it is evident that no matter how large the quantum number becomes, the
quantal distribution continues to oscillate ever more rapidly (although their vertical scale
stays finite). A perfect position detector would measure these oscillations,
so they must be smeared out in some way for the classical distribution to emerge. A general
analysis of the smearing process seems prohibitively difficult, and it is not entirely clear on
what principles it should be based.

In this paper we investigate this question based on Bohm's version of the spin-$j$
Einstein-Podolsky-Rosen experiment. This is an ideal system to study since
Bell inequalities continue to be violated with undiminished range as
$j \to \infty$\,\cite{gm82}.
Thus, the oft-made statement that the $j \to \infty$ limit corresponds to classical mechanics
needs to be examined more closely. The uncertainty principle
limits the precision with which {\it pairs\/} of noncommuting variables can be measured (or even
assigned values in an ontological sense) simultaneously. The nonclassicality revealed by the violation
of Bell inequalities suggests a stronger position, namely, that one must limit the absolute precision
with which an {\it individual\/} physical quantity can be measured. In other words, finite
precision of measurement should be regarded not just as an unavoidable fact of life, but as an
intrinsic ingredient of the classical limit.
This is an interesting shift in perspective, for one of the long-standing beliefs of
the classical mechanical world view has been that physical quantities can be measured to arbitrary
precision.

We propose in particular that any detector error protocol that is to be regarded as irreducible or
intrinsic to the classical limit should satisfy two principles: {\it agnosticism\/} and
{\it minimalism\/}. By agnosticism we mean that one should not be able to say whether the errors
arise in the detection process or in the state preparation process. And by minimalism we mean
that they should be no more than is needed to wash out the nonclassical features. Indeed, from this
point of view one should speak rather in terms of coarse-graining or smoothing-out ideal quantum
mechanical distributions than of error, although for brevity it is convenient to keep doing so.
We shall display a protocol that obeys both criteria, and we have also found protocols that disobey
one of the two.

In the spin-$j$ Einstein-Podolsky-Rosen-Bohm experiment, two particles of spin $j$ in the singlet
state, $\ket{\phi}$, fly toward two far apart detectors. The spin of one particle is measured
along a direction $\ahat_1$ and of the other along $\ahat_2$, with outcomes denoted $m_1$ and $m_2$.
The probability distribution for these outcomes is
$p_{\ahat_1\ahat_2}(m_1,m_2) = |\olap{\phi}{m_1 m_2}_{\ahat_1\ahat_2}|^2$, where
$\ket{m_1 m_2}_{\ahat_1\ahat_2} = \ket{j, m_1}_{\ahat_1} \otimes \ket{j, m_2}_{\ahat_2}$ is the simultaneous
eigenstate of $\bJ_1\cdot\ahat_1$ and $\bJ_2\cdot\ahat_2$ with eigenvalues $m_1$ and $m_2$.
Mermin and Schwarz (MS)\,\cite{ms82} discovered the pseudo-factorizable form, 
\beq
p_{\ahat_1\ahat_2}(m_1, m_2)
  = \int \frac{d^2\nhat}{4\pi} \, p_{\ahat_1} (m_1|\nhat) \, p_{\ahat_2}(m_2|-\nhat).
 \label{ms_factor_1}
\eeq
The one-axis functions $p_{\ahat_i}(m_i|\nhat)$ resemble conditional distributions for outcomes
$m_1$ and $m_2$ given a particular value for the hidden variable $\nhat$, which is a unit vector
that can point in any direction equiprobably. Since $p_{\ahat_1\ahat_2}(m_1, m_2)$ violates
Bell inequalities, the one-axis functions cannot be nonnegative, and they are not. MS showed that
\beq
p_{\ahat_i}(m|\nhat)
   = \frac{1}{d_j} \sum_{\ell = 0}^{2j}
        \sqrt{2\ell +1} f^j_{\ell}(m) P_{\ell}(\ahat_i\cdot\nhat),
   \label{p1_ms}
\eeq
where $d_j = 2j+1$, and $f^j_{\ell}(m)$ is an eigenvector of the real symmetric matrix
\beq
F_{mm'}(\tta) = | _{\zhat}\bra{j,m}{e^{i\tta J_y}}\ket{j,m'}_{\zhat}|^2,  \label{def_F_mat}
\eeq
the eigenvalue being the Legendre polynomial $P_{\ell}(\cos\tta)$. The functions
$f^j_{\ell}(x)$ turn out to be shifted and scaled discrete Chebyshev polynomials
[see \cite{szego}, or Ref.~\cite{ab_steg}, Sec.~22.17.], that are orthogonal with respect to the weight
function $\sum_{m=-j}^j \dta(x - m)$. Thus, there are only $2j+1$ of them with degree
$0 \le \ell \le 2j$. Either as polynomials, or as eigenvectors of $F_{mm'}$,
they obey orthonormality and completeness relations,
$d_j^{-1} \sum_{m=-j}^j f^j_{\ell}(m) f^j_{\ell'}(m) = \dta_{\ell \ell'}$,
$d_j^{-1} \sum_{\ell = 0}^{2j} f^j_{\ell}(m) f^j_{\ell}(m') = \dta_{mm'}$,
which fix them uniquely with the convention $f^j_{\ell}(j) > 0$. As special cases, we have
$f^j_0(m) = 1$, and $f^j_1(m) = [3/j(j+1)]^{1/2} m$.

A quantum state can also be characterized by its Wigner function, which is part of the
Weyl-Wigner-Moyal formalism\,\cite{weyl,wig,moy}, and which for spin may be implemented as
follows\,\cite{strato,kutzner,agarwal,sar78,vgb89,lbg13}. Any spin operator may be expanded in
terms of the complete set of spherical harmonic tensor operators, $\Ylm(\bJ)$, whose
Q, P, and Weyl symbols, denoted $\Phi^{Q,P,W}_{\ell m}(\nhat)$, are given by coefficients
$a^{Q,\, P,\, W}_{j \ell}$ times $\ylm(\nhat)$, where\,\cite{lbg13}
$a^Q_{j\ell} = \prod_{k=0}^{\ell}(j +\tshf - \tshf k)$,
$a^P_{j\ell} = \prod_{k=0}^{\ell}(j +\tshf + \tshf k)$,
and 
$a^W_{j \ell} = \bigl(a^P_{j \ell} a^Q_{j \ell} \bigr)^{1/2}$. (All $a_{j0}$'s are 1.)
Further,
$\Tr\bigl(\Ylm {\cal Y}^{\dagger}_{\ell'm'} \bigr)
    = (d_j/4\pi) \bigl(a^W_{j\ell}\bigr)^2 \dta_{\ell\ell'} \dta_{mm'}$.
These results enable us to construct and go between Q, P, and Weyl representations\,\cite{jens_me}.

The Wigner function for any system is nothing but the normalized Weyl symbol of the density matrix.
It purports, but often fails to be, a joint probability density for noncommuting phase space variables,
as it is not nonnegative. For our two spin system, the Wigner function, $W_{\rho}(\nhat_1, \nhat_2)$,
is nominally the probability that the spins point along $\nhat_1$ and $\nhat_2$. For any operator $G$,
$\avg{G} = \Tr(\rho G)
         = \int\!\!\int W_{\rho}(\nhat_1,\nhat_2) \Phi^W_G(\nhat_1,\nhat_2)\, d^2\nhat_1 d^2\nhat_2$,
where $\Phi^W_G$ is the Weyl symbol for $G$. (Putting $G = \Phi^W_G = 1$ gives the normalization of $W_{\rho}$.)
Hence, for the singlet state,
\beq
p_{\ahat_1 \ahat_2}(m_1, m_2)
  = \dint d^2\nhat_1\, d^2\nhat_2\,
     W_{\rho}(\nhat_1, \nhat_2) s^W(\nhat_1|m_1,\ahat_1) s^W(\nhat_2|m_2,\ahat_2),
  \label{epr_wigner_1}
\eeq
where $s^W(\nhat|m,\ahat)$ is the Weyl symbol of the projector
$\ket{m}_{\ahat}{_{\ahat}\bra{m}}$. We shall show below that
\beq
W_{\rho}(\nhat_1,\nhat_2)
  = \frac{1}{4\pi} 
     \sum_{\ell = 0}^{2j}\sum_{m=-\ell}^{\ell}
              \ylm(\nhat_1) \ylm^*(-\nhat_2).
\label{rho_W_yy}
\eeq
This is appealing in that as $j \to \infty$, the completeness sum of the spherical harmonics
yields $\dta(\nhat_1 + \nhat_2)/4\pi$, exactly as one would expect for an isotropic state of
two classical spinning gyroscopes with net angular momentum zero. The approach to this limit is
very singular, however. Using the addition theorem for the $\ylm$'s and the Christoffel-Darboux
theorem [see Ref.~\cite{szego}, Sec.~3.2, or Ref.~\cite{ab_steg}, Eq.~22.12.1], we find
(with $x = -\nhat_1\cdot\nhat_2$)
\beq
W_{\rho}
  = \frac{1}{(4\pi)^2} \frac{d_j}{1-x} [P_{2j}(x) - P_{2j+1}(x)].
\eeq
We plot $W_{\rho}$ in \fno{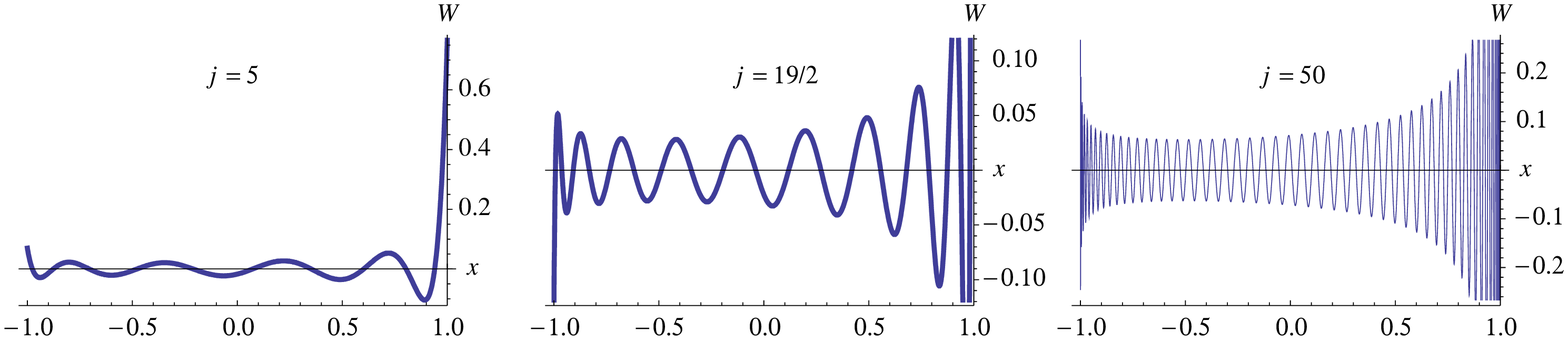}. That $W_{\rho} < 0$ is of course the standard
deficiency of the Wigner function, but as the plots show, as $j$ gets large, the oscillations
in $W_{\rho}$ get deeper and rapider\,\cite{W_osc}. In this way too we see that the singlet
does not get more classical as $j \to \infty$.
We show enlarged views of $W_{\rho}$ near $x = 1$ for $j=50$ in \fno{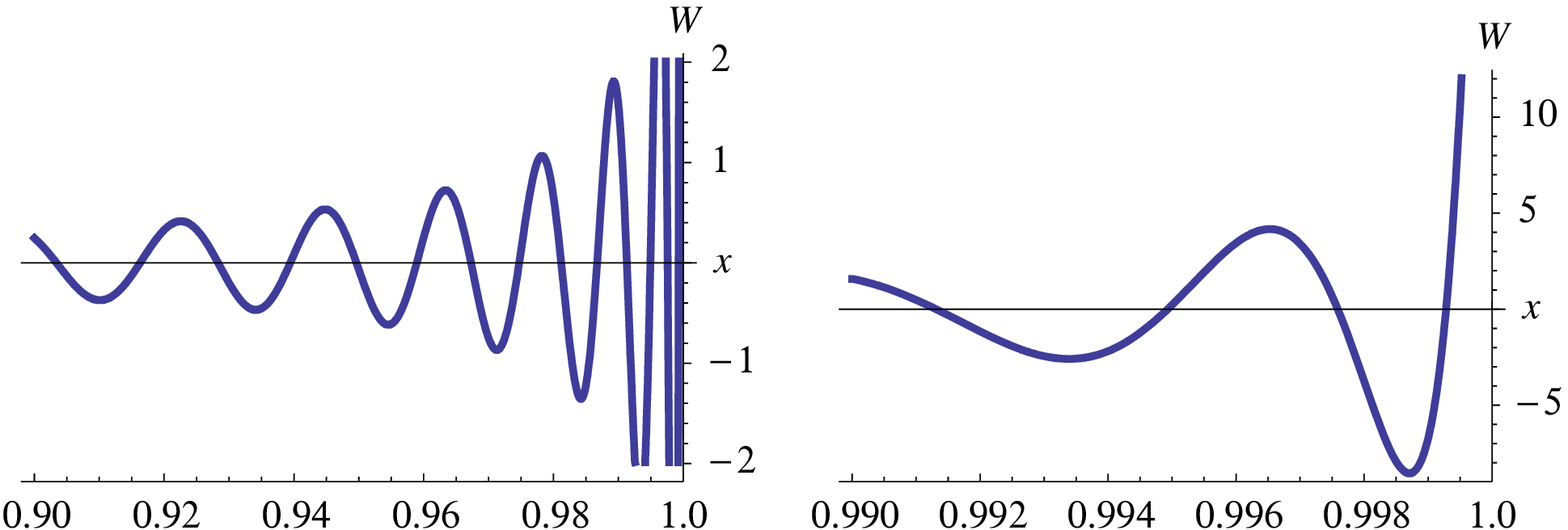}.

\begin{figure}[h]
\centering
\includegraphics[width=6in]{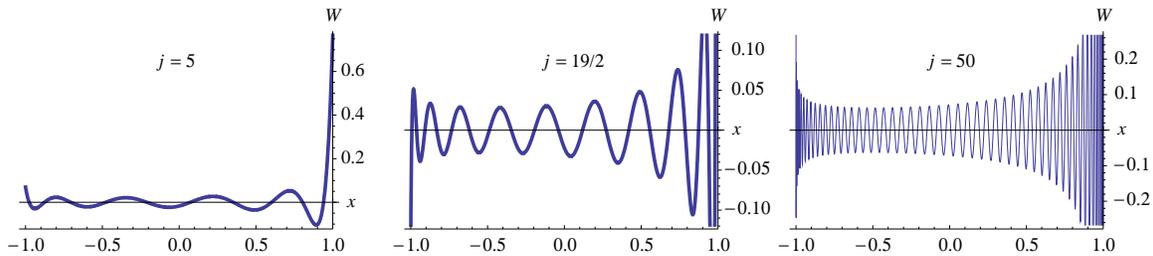}
\vskip0.1cm
\caption{Wigner functions for $j=5$, $19/2$, and $50$, as a function of $x = -\nhat_1\cdot\nhat_2$.
         The scales on the \ity axes should be noted. For $j=19/2$ and $50$, large portions of the vertical range
         are not shown.}
\label{wig_fns_epr.eps}
\end{figure}

\begin{figure}[h]
\centering
\includegraphics[width=4in]{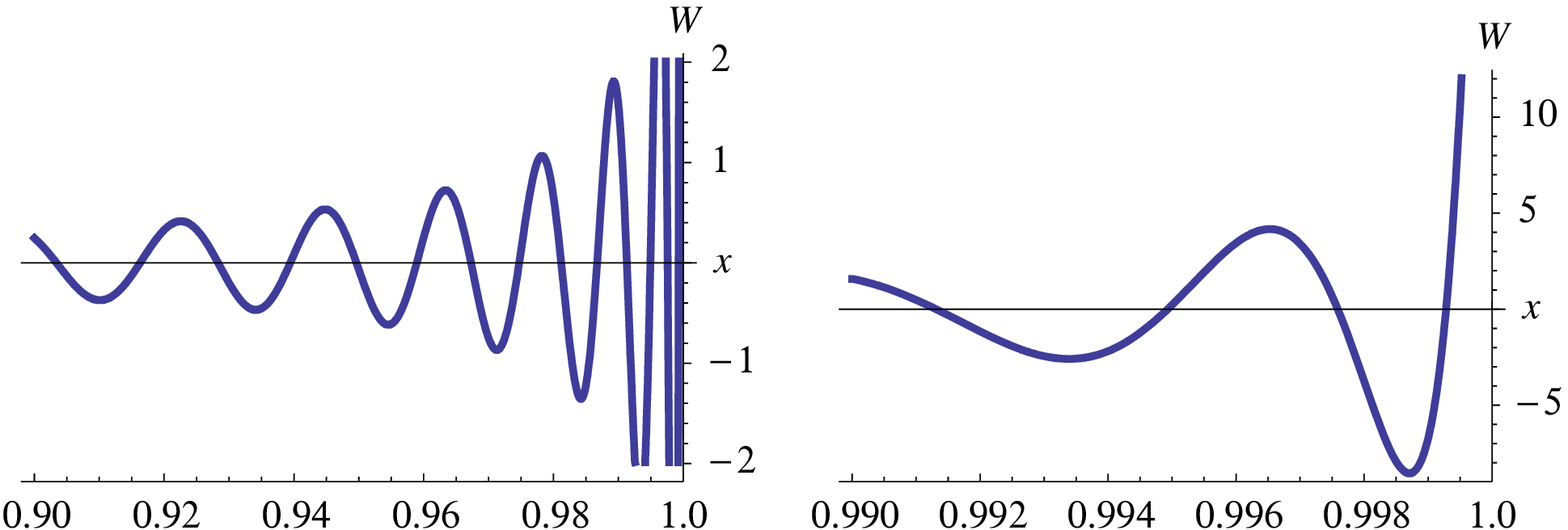}
\vskip0.1cm
\caption{Enlarged view near $x = 1$ of the Wigner function for $j=50$. Once again, the
         scales on the \ity axes should be noted.}
\label{wig_epr_enl.eps}
\end{figure}

To show \eno{rho_W_yy}, and for other technical parts of our analysis, we shall rely repeatedly on
two facts. First is the spectral representation of $F_{mm'}$,
\beq
F_{mm'}(\tta)
  = \frac{1}{d_j} \sum_{\ell = 0}^{2j}
       f^j_{\ell}(m) f^j_{\ell}(m') 
          P_{\ell}(\cos\tta).
  \label{F_spectral}
\eeq
Second is a connection between the polynomials $f^j_{\ell}$ and the Weyl-to-Q map,
\beq
f^j_{\ell}(j) = \sqrt{2\ell + 1} a^Q_{j\ell}/a^W_{j\ell},  \label{flj_answer}
\eeq
which can be proved by induction.

To find $W_{\rho}$ for the singlet, we first find the Q-symbol, $\Phi^Q_{\rho}$. 
With $\ket{\nhat}$ being a spin coherent state\,\cite{radcliffe,arecchi}, we have
$\Phi^Q_{\rho}(\nhat_1,\nhat_2) = \bigl|\bra{\phi}(\ket{\nhat_1} \otimes \ket{\nhat_2}) \bigr|^2
   = d_j^{-1}|\olap{-\nhat_1}{\nhat_2}|^2
   = d_j^{-1} F_{jj}(\gam)$,
with $\cos\gam = -\nhat_1\cdot\nhat_2$. Next, we use the spectral representation (\ref{F_spectral}),
and multiply each term in the sum over $\ell$ by $(a^W_{j\ell}/a^Q_{j\ell})^2$ to convert to the
Weyl symbol. Last, we multiply by $(d_j/4\pi)^2$ for normalization, and employ the
the connection (\ref{flj_answer}). This leads directly to \eno{rho_W_yy}.

We include detector error via a matrix $R$, such
that $R_{mm'}$ is the probability that a particle which has spin $m'$ is detected in the
bin for spin $m$. Since particles can neither be lost nor appear from nowhere, we must have
$\sum_m R_{mm'} = \sum_{m'} R_{mm'} = 1$,
and to be probabilities, we must have $R_{mm'} \ge 0$ for all $m$, $m'$. Using the same
error matrix at both detectors, the measured distribution 
${\bap}_{\ahat_1\ahat_2}(m_1, m_2)$ is given by the same form as \eno{ms_factor_1} with the
one-axis functions replaced by
\beq
\bap_{\ahat_i}(m|\nhat) 
  = \sum_{m'} R_{mm'} \, p_{\ahat_i}(m'|\nhat).
\eeq
We will say that the error matrix is {\it sufficient\/}
if $\bap_{\ahat_i}(m|\nhat) \ge 0$ for all $m$, $\nhat$, and $\ahat_i$, and {\it minimally
sufficient\/} if
the value 0 is attained for some set of parameters. The smoothed one-axis functions then have
meaning as conditional probability distributions, and $\bap_{\ahat_1\ahat_2}(m_1 m_2)$ is rendered
classical in that it cannot violate any Bell inequalities.

In terms of the Wigner function approach, the error matrices effect transformations on the
Weyl symbols $s^W \to {\bas^W}$
[$\bas^W(\nhat|m,\ahat) = \sum_{m'} R_{mm'} s^W(\nhat|m',\ahat)$],
which will generally not be equivalent to a transformation of $W_{\rho}$. When it is, the
error protocol is agnostic. To see when this is possible, we must examine $s^W$.

To find $s^W$, we proceed as we did for $W_{\rho}$: find the Q symbol of
$\ket{m}_{\ahat}{_{\ahat}\bra{m}}$, expand in the $\ylm$'s, and multiply each term by
$a^W_{j\ell}/a^Q_{j\ell}$. Now,
$s^Q(\nhat|m,\ahat) = |\olap{\nhat}{m,\ahat}|^2 = F_{jm}(\al)$,
where $\cos \al = \ahat\cdot \nhat$. Using the spectral representation (\ref{F_spectral})
and the connection (\ref{flj_answer}), we obtain
\beq
s^W(\nhat|m,\ahat) = \frac{1}{d_j} \sum_{\ell} \sqrt{2\ell + 1} f^{\ell}_m P_{\ell}(\nhat\cdot\ahat).
  \label{single_W}
\eeq
That is to say, $s^W(\nhat|m,\ahat) \equiv p_{\ahat}(m|\nhat)$. The MS one-axis functions are the Weyl
symbols for projection operators onto the states $\ket{j,m}_{\ahat}$.

The equivalence just noted leads us to consider error matrices that are
eigenoperators for the $f^j_{\ell}(m)$, i.e.,
\beq
\sum_{m'} R_{mm'} f^j_{\ell}(m') = c_{\ell} f^j_{\ell}(m),
  \label{R_eigen}
\eeq
where the $c_{\ell}$ are arbitrary except that $c_0 = 1$. Then,
$\bap_{\ahat}(m|\nhat) 
  = {\bas}^W(\nhat|m,\ahat)
  = d_j^{-1} \sum_{\ell} \sqrt{2\ell + 1} c_{\ell} f^{\ell}_m P_{\ell}(\nhat\cdot\ahat)$.
Because of the orthogonality of the Legendre polynomials, the factors $c_{\ell}$ can
be transferred onto the Wigner function. That is, $\bap_{\ahat_1\ahat_2}$ can be put in
the form (\ref{rho_W_yy}), where we leave the single spin Weyl symbols, the $s^W$'s, untouched,
and replace $W_{\rho}$ by
\beq
W_{\bar\rho}(\nhat_1,\nhat_2)
  = \frac{1}{4\pi} 
     \sum_{\ell = 0}^{2j}\sum_{m=-\ell}^{\ell}
              c_{\ell}^2 \ylm(\nhat_1) \ylm^*(-\nhat_2).  \label{rho_Wbar_yy}
\eeq
But this says that the error or coarse-graining protocol affects the state preparation,
and not the detection, i.e., is agnostic. Hence, we have shown that error matrices obeying
\eno{R_eigen} are agnostic. (They also preserve the isotropy of the singlet state.)

An agnostic but trivial protocol is obtained by taking $c_{\ell} = \dta_{\ell 0}$, i.e.,
$R_{mm'} = d_j^{-1}$ for all $m$, $m'$. Now $\bap_{\ahat_1\ahat_2}(m_1 m_2) = d_j^{-2}$,
and the two spins are totally uncorrelated. Hence agnosticism by itself is not a compelling
principle, and we must consider sufficiency also.

The special error protocol that we have found that is both agnostic and minimal arises from
choosing
\beq
c_{\ell} = \frac{a^Q_{j\ell}}{a^W_{j\ell}}
  = \frac{f^j_{\ell}(j)}{\sqrt{2\ell + 1}}
  = \prod_{k=0}^{\ell} \Bigl( \frac{2j + 1 -k}{2j + 1 + k} \Bigr)^{1/2}.
  \label{c_ell_I}
\eeq
Then, with $\cos\al = \nhat\cdot\ahat$ and $\cos\gam = -\nhat_1\cdot\nhat_2$,
\beq
{\bap}_{\ahat}(m|\nhat)
   = F_{jm}(\al)
   = \binom{2j}{j-m}
    \bigl[\tshf(1 + \cos\al) \bigr]^{j+m}
    \bigl[\tshf(1 - \cos\al) \bigr]^{j-m},
  \label{good_1_axis}
\eeq
and
\beq
W_{\bar\rho}(\nhat_1,\nhat_2)
  = \frac{d_j}{(4\pi)^2} F_{jj}(\gam)
  = \frac{d_j}{(4\pi)^2} \Bigl(\frac{1 + \cos\gam}{2} \Bigr)^{2j},
  \label{good_wigner}
\eeq
both of which are nonnegative.
This protocol is minimal because the $\bap_{\ahat_i}(m_i|\nhat)$ do in fact vanish when
$\nhat = \ahat_i$ for any $m_i < j$, and when $\nhat = -\ahat_i$ for any $m_i > -j$. These
one-axis functions provide an explicit locally realistic model for the resulting
coarse-grained distribution $\bap_{\ahat_1\ahat_2}(m_1 m_2)$. It is also minimal from the
point of view of state preparation, because $W_{\bar\rho}(\nhat_1,\nhat_2)$ does vanish
when $\nhat_1 = \nhat_2$.

Completeness of the $f^j_{\ell}(m)$ allows any error matrix that obeys \eno{R_eigen}
to be written as
\beq
R_{mm'} = \frac{1}{d_j} \sum_{\ell} c_{\ell} f^j_{\ell}(m) f^j_{\ell}(m').
\eeq
(The constraint $c_0 = 1$ follows from $f^j_0(m) = 1$ and the demands
$\sum_m R_{mm'} = \sum_{m'} R_{mm'} = 1$.) In addition, we must have $R_{mm'} \ge 0$ for
all $m$, $m'$. We have not been able to establish rigorously that this is so for our
special protocol (\ref{c_ell_I}), but we have verified it by hand for $d_j \le 4$, and
numerically for $d_j \ltwid 20$, beyond which our numerical precision is not sufficient.
A crude argument is as follows.
\eno{c_ell_I} implies
$R_{mm'} = \hf d_j \int_{-1}^1 dx\, h^j_m(x) F_{jm'}(\cos^{-1}x)$,
with
$h^j_m(x) = d_j^{-1} \sum_{\ell = 0}^{2j} \sqrt{2\ell + 1} f^j_{\ell}(m) P_{\ell}(x)$.
As $j \to \infty$, $F_{jm'}$ and $h^j_m(x)$ are highly peaked functions of $x$ with maxima
at $x_{m'} = m'/j$ and $x_m = m/j$, and both of them integrate to $2/d_j$. $F_{jm'}$ is a binomial
distribution, which is like a Gaussian of width $\sig_x = (1 - x^2_{m'}/2j)^{1/2}$
(for $m'$ sufficiently far away from $\pm j$). The sum for $h^j_m(x)$ resembles a
Christoffel-Darboux sum, which suggests that for large $j$, $h^j_m(x)$ is much more narrowly
peaked with a width of order $1/j^2$. Away from its maximum, it oscillates on a scale
$j^{1/2}$ with an approximate frequency $1/2j$. This frequency and the width are both much
smaller than the width of the Gaussian, so $h^j_m(x)$ effectively behaves as
$d_j^{-1}\dta(x-x_m)$ in the integral. Hence, for $m$, $m'$ not too close to $\pm j$,
\beq
R_{mm'}
  \simeq  \frac{1}{j\sqrt{2\pi\sig_x^2}} e^{-(x_m - x_{m'})^2/2\sig_x^2}.
 \label{R_I_gaussian}
\eeq
This says that as $j \to \infty$, the interval $|m-m'|/j$ over which coarse graining is
necessary becomes of order $j^{-1/2}$. 

It is not easy to come up with a simple figure of merit for the sufficiency of error
matrices for $j > 1/2$, although one choice is to see by how much the spin
correlation is reduced. For protocols obeying \eno{R_eigen}, this factor is easily shown to
be $c_1^2$, so that for our special protocol (\ref{c_ell_I}),
$\avg{(\bJ_1\cdot\ahat) (\bJ_2\cdot\bhat)} = j^2\ahat\cdot\bhat/3$. However, for $j=1/2$, where
the detector error reduces to a single number, it is known not to be the least conservative
smoothing procedure. It corresponds to an error rate of 21.1\%\,\cite{ms82}. In Ref.~\cite{ag83}
we found a truly positive factorizable representation for $p_{\ahat_1\ahat_2}(m_1 m_2)$ based
on an error rate of 14.6\%. Whether this number can be further reduced for $j=1/2$, or whether
there are analogous constructions for higher $j$ that are less conservative than the protocol
(\ref{c_ell_I}) remain open questions.

We conclude by mentioning two other error protocols for comparison.
In Ref.~\cite{ag_thesis}, the author considered
\beq
R_{mm'} = \hf d_j \int_{x_{m-}}^{x_{m+}} d(\cos\tta) F_{jm'}(\tta),
 \label{R_thesis}
\eeq
where
$x_{m\pm} = (2m \pm 1)/d_j$.
Now $R_{mm'} > 0$, but this protocol is not agnostic. For small $j$ we can explicitly show
it is insufficient. The insufficiency decreases for larger $j$, so it is very likely that this
defect could be repaired by admixing in the trivial protocol $R_{mm'} = d_j^{-1}$ in an amount
that decreases steadily as $j \to \infty$. This yields a gnostic but sufficient protocol.
An agnostic but oversufficient protocol is obtained if we choose
$c_{\ell} = \bigl(a^Q_{j\ell}/a^W_{j\ell} \bigr)^2$. In this case, we can prove that
$R_{mm'} > 0$. 

I am indebted to N.~D. Mermin for an old collaboration in which we studied similar questions,
and specifically for suggesting the form (\ref{R_thesis}).


\begin{thebibliography}{99}
\bibitem{W_osc}
$(4\pi)^2 W_{\rho}$ equals $(2j+1)^2$ at $x=1$, $(-1)^j (2j+1)$ at $x=-1$, the scale of
the oscillations is $\sim j^{1/2}$, and the first zero of $W_{\rho}$ is at a distance
of $7.34/(2j+1)^2$ from the main peak at $x = 1$. These results, and asymptotic forms for
$W_{\rho}$ will be published elsewhere.
\end{thebibliography}
\end{document}